\documentstyle[12pt,epsfig,axodraw]{article}

\advance\textheight by \topskip
\begin{document}
\tolerance=100000
\thispagestyle{empty}
\setcounter{page}{1}

\newcommand{\HPA}[1]{{\it Helv.\ Phys.\ Acta.\ }{\bf #1}}
\newcommand{\AP}[1]{{\it Ann.\ Phys.\ }{\bf #1}}
\newcommand{\be}{\begin{equation}}
\newcommand{\ee}{\end{equation}}
\newcommand{\br}{\begin{eqnarray}}
\newcommand{\er}{\end{eqnarray}}
\newcommand{\ba}{\begin{array}}
\newcommand{\ea}{\end{array}}
\newcommand{\bi}{\begin{itemize}}
\newcommand{\ei}{\end{itemize}}
\newcommand{\bn}{\begin{enumerate}}
\newcommand{\en}{\end{enumerate}}
\newcommand{\bc}{\begin{center}}
\newcommand{\ec}{\end{center}}
\newcommand{\ul}{\underline}
\newcommand{\ol}{\overline}
\def\l{\left\langle}
\def\r{\right\rangle}
\def\as{\alpha_{s}}
\def\ycut{y_{\mbox{\tiny cut}}}
\def\yij{y_{ij}}
\def\epem{\ifmmode{e^+ e^-} \else{$e^+ e^-$} \fi}
\newcommand{\eeww}{$e^+e^-\rightarrow W^+ W^-$}
\newcommand{\qqQQ}{$q_1\bar q_2 Q_3\bar Q_4$}
\newcommand{\eeqqQQ}{$e^+e^-\rightarrow q_1\bar q_2 Q_3\bar Q_4$}
\newcommand{\eewwqqqq}{$e^+e^-\rightarrow W^+ W^-\ar q\bar q Q\bar Q$}
\newcommand{\eeqqgg}{$e^+e^-\rightarrow q\bar q gg$}
\newcommand{\eeqloop}{$e^+e^-\rightarrow q\bar q gg$ via loop of quarks}
\newcommand{\eeqqqq}{$e^+e^-\rightarrow q\bar q Q\bar Q$}
\newcommand{\eewwjjjj}{$e^+e^-\rightarrow W^+ W^-\rightarrow 4~{\rm{jet}}$}
\newcommand{\eeqqggjjjj}{$e^+e^-\rightarrow q\bar 
q gg\rightarrow 4~{\rm{jet}}$}
\newcommand{\eeqloopjjjj}{$e^+e^-\rightarrow q\bar 
q gg\rightarrow 4~{\rm{jet}}$ via loop of quarks}
\newcommand{\eeqqqqjjjj}{$e^+e^-\rightarrow q\bar q Q\bar Q\rightarrow
4~{\rm{jet}}$}
\newcommand{\eejjjj}{$e^+e^-\rightarrow 4~{\rm{jet}}$}
\newcommand{\jjjj}{$4~{\rm{jet}}$}
\newcommand{\qqbar}{$q\bar q$}
\newcommand{\ww}{$W^+W^-$}
\newcommand{\ar}{\rightarrow}
\newcommand{\sm}{${\cal {SM}}$}
\newcommand{\Dir}{\kern -6.4pt\Big{/}}
\newcommand{\Dirin}{\kern -10.4pt\Big{/}\kern 4.4pt}
\newcommand{\DDir}{\kern -8.0pt\Big{/}}
\newcommand{\DGir}{\kern -6.0pt\Big{/}}
\newcommand{\wwqqqq}{$W^+ W^-\ar q\bar q Q\bar Q$}
\newcommand{\qqgg}{$q\bar q gg$}
\newcommand{\qloop}{$q\bar q gg$ via loop of quarks}
\newcommand{\qqqq}{$q\bar q Q\bar Q$}

\def\st{\sigma_{\mbox{\scriptsize t}}}
\def\Ord{\buildrel{\scriptscriptstyle <}\over{\scriptscriptstyle\sim}}
\def\OOrd{\buildrel{\scriptscriptstyle >}\over{\scriptscriptstyle\sim}}
\def\pl #1 #2 #3 {{\it Phys.~Lett.} {\bf#1} (#2) #3}
\def\np #1 #2 #3 {{\it Nucl.~Phys.} {\bf#1} (#2) #3}
\def\zp #1 #2 #3 {{\it Z.~Phys.} {\bf#1} (#2) #3}
\def\jp #1 #2 #3 {{\it J.~Phys.} {\bf#1} (#2) #3}
\def\pr #1 #2 #3 {{\it Phys.~Rev.} {\bf#1} (#2) #3}
\def\prep #1 #2 #3 {{\it Phys.~Rep.} {\bf#1} (#2) #3}
\def\prl #1 #2 #3 {{\it Phys.~Rev.~Lett.} {\bf#1} (#2) #3}
\def\mpl #1 #2 #3 {{\it Mod.~Phys.~Lett.} {\bf#1} (#2) #3}
\def\rmp #1 #2 #3 {{\it Rev. Mod. Phys.} {\bf#1} (#2) #3}
\def\cpc #1 #2 #3 {{\it Comp. Phys. Commun.} {\bf#1} (#2) #3}
\def\sjnp #1 #2 #3 {{\it Sov. J. Nucl. Phys.} {\bf#1} (#2) #3}
\def\xx #1 #2 #3 {{\bf#1}, (#2) #3}
\def\hepph #1 {{\tt hep-ph/#1}}
\def\preprint{{\it preprint}}

\begin{flushleft}
{RAL-TR-99-070}
\end{flushleft}
\begin{flushright}
\vskip-1.0truecm
{LC-TH-1999-017\\
November 1999}
\end{flushright}
\vskip0.1cm\noindent
\begin{center}
{\Large {\bf Six-jet production at $e^+e^-$ linear 
colliders\footnote{Talk given at the 2nd ECFA/DESY Study on Physics 
and Detectors for a Linear Electron-Positron Collider, Lund,
Sweden, 28-30 June 1998.}}}\\[0.5cm]
{\large 
S.~Moretti}\\[0.05 cm]
{\it Rutherford Appleton Laboratory, Chilton, Didcot, Oxon OX11 0QX, UK.}
\end{center}

\begin{abstract}
{\small
\noindent
The calculation of the tree-level QCD processes
$e^+e^-\ar q\bar q gggg$, 
$q\bar q q'\bar q' gg$ and
$q\bar q q'\bar q' q''\bar q''$ has recently been accomplished. 
We highlight here the relevance of such reactions for some of
the physics at future electron-positron linear accelerators.}
\end{abstract}
\vskip0.25cm\noindent
{\large\bf 1. Motivations for an exact calculation of $e^+e^-\to 6$ partons}
\vskip0.15cm\noindent
As accelerator physics will enter the Linear Collider (LC) epoch 
\cite{ee500,newee500}, 
one will encounter a long series of resonant processes ending up 
with six-jet signatures.
One should recall top quark production and decay for a start, whose
study will
represent  one of the main areas of activity at a future LC \cite{top}. 
Top quarks will be produced in pairs, via $e^+e^-\to \gamma^*,Z^*\to t\bar t$,
followed most of the times by $t\bar t\ar b\bar b W^+W^-\ar \mbox{6~jets}$.
Then one should not 
forget the new generation of gauge boson resonances, such as 
$e^+e^-\to ZW^+W^-$ and
$ZZZ$, and their dominant six-jet decays. The interest in these reactions
resides 
primarily in the possibility of an accurate study of the gauge structure
of the electroweak (EW) model
\cite{vectorbosons}. In the same respect, one 
could also add highly-virtual photonic processes, 
like $e^+e^-\to\gamma^* W^+W^-$, 
$\gamma^* ZZ$, $\gamma^*\gamma^* Z$ and $\gamma^*\gamma^*\gamma^*$, in
which the photons split into quark-antiquark pairs. 
In addition,  of particular relevance are reactions involving the 
Higgs particle, $\phi$, e.g., in the Standard Model (SM),
 such as $e^+e^-\to Z\phi\ar ZW^+W^-$ and $Z\phi\ar ZZZ$ -- as discovery
channels of a heavy scalar boson \cite{higgs} -- or $e^+e^-\to
Z\phi\ar Z\phi\phi$ -- as
a means to study the Higgs potential of a light scalar
(the latter decaying to $b\bar b$) \cite{triple}. 

Given such a wide scope offered by six-jet final states, 
it is of paramount importance to have a strong control on the 
backgrounds.
The parton-shower (PS) event generators (e.g.,
HERWIG \cite{HERWIG} and JETSET/PYTHIA \cite{JETSET})
 represent a valuable instrument in this respect, as they are able to 
describe the full event, from the initial hard scattering down to the hadron 
level. However, Matrix Element (ME) models are acknowledged to 
describe the large angle distributions of the QCD  radiation
better than the former  do (see, e.g., \cite{spin}), which are in 
fact superior in the small angle dynamics. As in the processes we just
mentioned the final state jets are typically produced at large angle
and are isolated (being  the decay products of massive objects), 
 the need of exact ME computations should be 
manifest\footnote{The matching of fixed-order (as well as resummed)
multi-parton final states with the subsequent PS to 
finally reach the hadronisation
stage is also a pressing matter, towards which some
progress has recently been made \cite{matching}.}.
As for theoretical advances in this respect, 
studies of $e^+e^-\ar$~6-quark EW processes are well under way
(see Ref.~\cite{nico5} for a review). However, a large fraction 
of the six-jet cross section  comes
from QCD interactions. The case of QCD six-jet production from $W^+W^-$ decays
was considered in Ref.~\cite{WW6}.
In this note, we discuss the dominant, tree-level QCD contributions to
six-jet final states through  the order ${\cal O}(\as^4)$, i.e., the processes:
\be\label{sixparton}
e^+e^-\ar \gamma^*,Z^*\to q\bar q gggg,~ 
                          q\bar q q'\bar q' gg,~
                          q\bar q q'\bar q' q''\bar q'',
\ee
where $q, q'$ and $q''$ represent any possible flavours of quarks
(massless and/or massive) and $g$ is a gluon, whose computation has
recently been tackled \cite{ee6j}. 
\vskip0.5cm\noindent
{\large\bf 2. Numerical results}
\vskip0.15cm\noindent
In order to select a six-`jet' sample we apply a jet 
clustering algorithm directly to the `quarks' and `gluons'
 in the final state of the processes (\ref{sixparton}). For 
illustrative purposes, we use the Cambridge (C) jet-finder 
\cite{cambridge,schemes} only. This is based 
on the `measure'
\begin{equation}\label{D}
y_{ij} = {{2\min (E^2_i, E^2_j)(1-\cos\theta_{ij})}\over{s_{ee}}},
\end{equation}
where $E_i$ and $E_j$ are the energies and $\theta_{ij}$ the separation
of any pair $ij$ of particles in the final state, with $i<j=2, ... 6$, to be 
compared against a resolution parameter denoted by $y$. 
In our tree-level approximation, the selected rate is nothing else 
than the total partonic
cross section with a cut $y_{ij}>y$ on any possible $ij$ 
combination. 
The summations over the three reactions (\ref{sixparton}) and 
over all possible `massless' (see Ref.~\cite{ee6j} for a dedicated
study of mass effects)
combinations of quark flavours in each of these have been performed here.
As for numerical inputs, they can be found in \cite{ee6j}.

\begin{figure}[!t]
\begin{center}
\vskip-0.75cm
~\epsfig{file=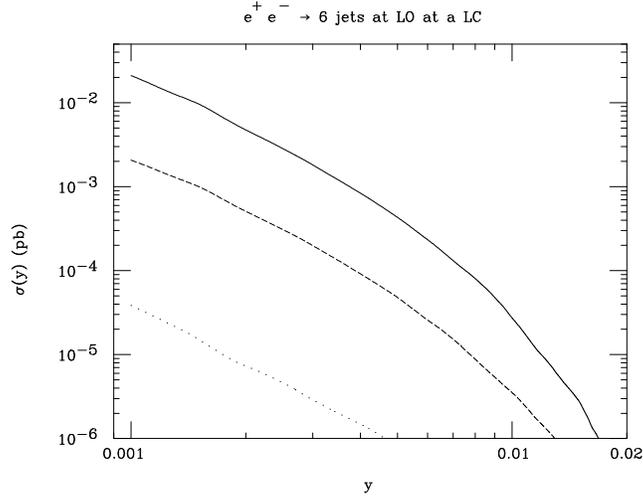,width=7.5cm,height=10cm,angle=90}
\vskip-1.0cm\caption{\small The total cross section of six-jet events 
at LO in the C scheme, decomposed in terms of the three contributions
$q\bar q gggg$ (solid), $q\bar q q'\bar q' gg$
(dashed) and $ q\bar q q'\bar q' q''\bar q''$ (dotted).}
\label{fig_comp6nlc}
\end{center}
\end{figure}

The six-jet event rate induced by the ${\cal O}(\as^4)$ QCD events at
$\sqrt{s_{ee}}=500$ GeV -- the value that we use here 
for the centre-of-mass (CM) energy of a LC --
can be rather large. Adopting a yearly
luminosity of, e.g., 100 fb$^{-1}$ and assuming a 
standard evolution of $\alpha_s$ with increasing energy, at 
$y=0.001$ one should expect some  
2300 events per annum: see Fig.~\ref{fig_comp6nlc}.
However, these rates decrease rapidly as $y$ 
 gets larger. 
From Fig.~\ref{fig_comp6nlc}, one can appreciated how the 
 dominant component is due to two-quark-four gluon events,
followed by the four-quark-two-gluon and six-quark ones, respectively. These
 relative rates are of particular
relevance to a LC environment. In fact, the capability of the detectors
of distinguishing between jets due to quarks (and among these, bottom flavours
in particular: e.g., in selecting top and light Higgs decays) 
and gluons, is of crucial importance there, in order to
perform dedicated searches for old and new particles. 

\begin{figure}[!b]
\begin{center}
\vskip-0.5cm
~\epsfig{file=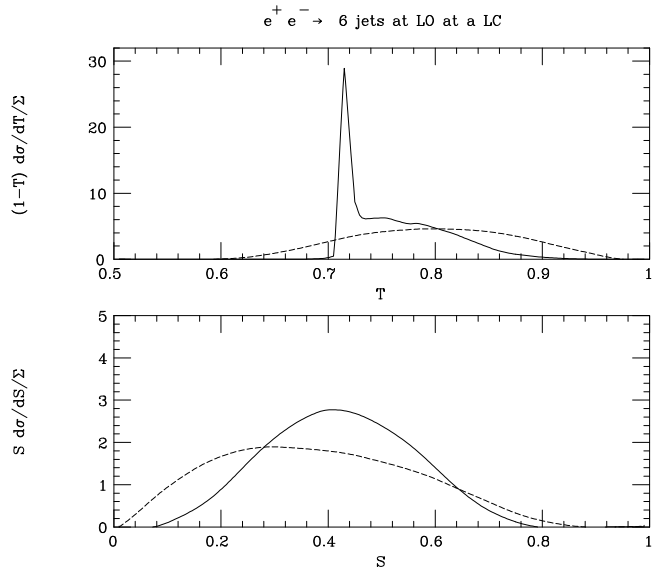,width=7.5cm,height=10cm,angle=90}
\vskip-0.35cm\caption{\small The distributions in thrust (upper plot) and
sphericity (lower plot) for $ t\bar t\ar6$-jet
events (solid) 
and for those of the  type (\ref{sixparton}) (dashed) at LO 
in the C scheme with $y=0.001$. 
Notice that the distributions have been normalised to a common
factor (one) for readability.}
\label{fig_shape_tt}
\end{center}
\end{figure}

The concern about background effects at a LC 
due to six-jet events via ${\cal O}(\alpha_s^4)$ QCD comes about if one 
considers that they  may naturally
survive some of the top signal selection 
criteria. It is well known that the
large value of the top mass (here, $m_t=175$ GeV) 
leads to rather spherical events. Therefore, shape variables such us 
thrust and sphericity 
represent useful means to disentangle $e^+e^-\ar t\bar t$ events.
For example, a selection strategy that does not exploit
neither lepton identification nor the tagging of $b$-jets was outlined 
in Ref.~\cite{top}.
The requirements are a large particle multiplicity, a high number of jets
(eventually forced to six) and a rather 
low(high)  thrust(sphericity). Jets are selected according to a jet clustering 
algorithm (in our
case, the C one with $y=0.001$, for sake of illustration).
Clearly, six-jet events (\ref{sixparton}) meet the first two criteria. 
As for the thrust and sphericity distributions, these are
 shown in Fig.~\ref{fig_shape_tt}.
From there, it is evident the overlap of the 
$t\bar t$ and QCD spectra.

\begin{figure}[!t]
\begin{center}
\vskip-3.0cm
~\epsfig{file=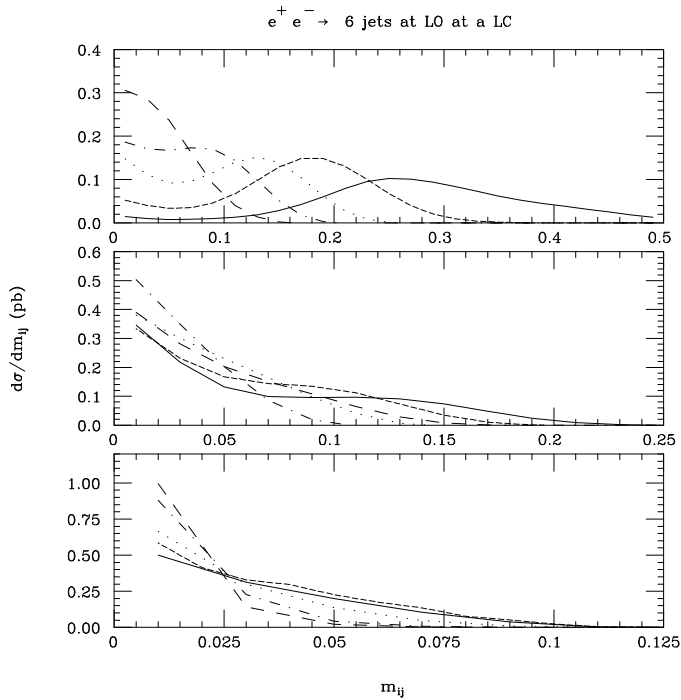,width=10cm,height=13.33cm,angle=0}
\vskip-2.0cm\caption{\small 
The distributions in the reduced invariant mass $m_{ij}$ (\ref{mij}) 
for events of the type (\ref{sixparton}) at LO in the C scheme
with $y=0.001$, for the following combinations of parton pairs $ij$: 
$({12})[{23}]\{{35}\}$ (solid),  
$({13})[{24}]\{{36}\}$ (short-dashed),  
$({14})[{25}]\{{45}\}$ (dotted),  
$({15})[{26}]\{{46}\}$ (dot-dashed) and  
$({16})[{34}]\{{56}\}$ (long-dashed),  
in the (upper)[central]\{lower\} frame.}
\label{fig_newmasses_6j_nlc}
\end{center}
\end{figure}

Searches for resonances will often need to 
rely on the mass reconstruction of di-jet systems. Therefore, it is 
worth looking at the invariant mass distributions
which will be produced  
by all possible two-parton combinations $ij$ in (\ref{sixparton}). 
As usual in multi-jet analyses,
we first order the jets in energy, so that $E_1>...>E_6$. Then,
we construct
\be
\label{mij}
m_{ij}\equiv\frac{M_{ij}^2}{s_{ee}}=\frac{2E_iE_j(1-\cos\theta_{ij})}{s_{ee}}.
\ee
These fifteen quantities 
are shown in Fig.~\ref{fig_newmasses_6j_nlc} for the C scheme at $y=0.001$.
We found it convenient to plot the `reduced' invariant masses $m_{ij}$ rather
than the actual ones $M_{ij}$, as energies and angles `scale' with the CM 
energy in such a way that the shape of the distributions is largely unaffected
by changes in the value of $\sqrt s_{ee}$ in the CM 
energy range relevant to a LC.
In the figure, it is interesting to notice the `resonant'
behaviour of some of the distributions. 

\begin{figure}[!h]
\begin{center}
\vskip-1.0cm
~\epsfig{file=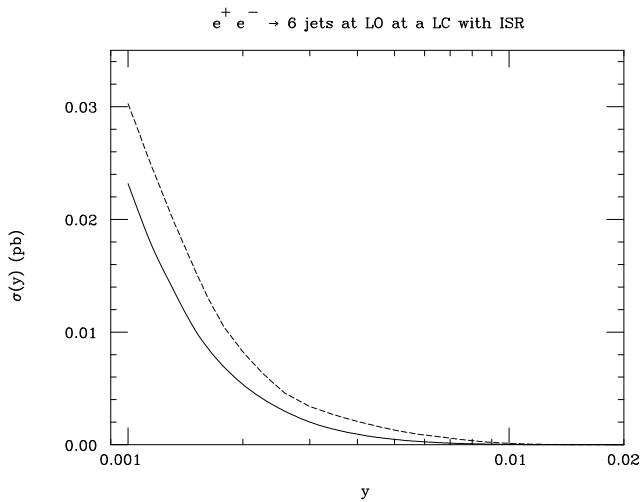,width=7.5cm,height=10cm,angle=90}
\vskip-1.0cm\caption{\small The total cross section of six-jet events
at LO in the D scheme, without (solid) and with (dashed) ISR.}
\label{fig_ISR}
\end{center}
\end{figure}

It is well known that photon bremsstrahlung  generated
by the incoming $e^+e^-$ beams 
(or Initial State Radiation, ISR)
can be quite sizable at a LC\footnote{Other peculiar feature is 
the presence
of Linac energy spread and beamsstrahlung \cite{bark} effects. For a
TESLA collider design, these are however smaller in comparison \cite{ISR},
so we leave them aside here.}.
In order to implement ISR we  resort to the 
so-called Electron Structure Function approach by
using   the  ${\cal O}(\alpha_{em}^2)$ expressions of
Ref.~\cite{Nicro}. As a simple exercise, we plot the production
rates for the sum of the processes (\ref{sixparton}) in presence
of ISR in Fig.~\ref{fig_ISR}, as a function of the resolution
parameter $y$ in the C scheme. They are compared to those obtained
without ISR (that is, the sum of the rates in Fig.~\ref{fig_comp6nlc}).
We see that the curve corresponding to the processes 
 convoluted with ISR lies above the
lowest-order one. This is rather intuitive,
as it is well known that the radiation of photons from the
incoming electron and positrons tends to lower the `effective' CM energy
of the collision \cite{ISR}.  At $y=0.001$, the difference in the rates
is around $25\%$ and this tends to increase as $y$ gets larger. 
In contrast, we have checked that the shape of the 
differential distributions (such as those 
in Fig.~\ref{fig_newmasses_6j_nlc}) suffers little from
ISR.

\vskip0.5cm\noindent
{\large\bf 3. Summary}
\vskip0.15cm\noindent
The exact calculation of the processes
$e^+e^-\ar q\bar q gggg$, $q\bar q q'\bar q' gg$ and  
$q\bar q q'\bar q' q''\bar q''$ (for both 
massless and massive quarks) at the leading-order in perturbative QCD
has been completed. 
In this short note we have emphasised the strong impact that
such reactions can have as backgrounds to many of the multi-jet studies 
foreseen at a LC. A numerical program is available for simulations,
based on helicity amplitudes, thus also allowing for initial state
polarisation.
\vskip0.15cm\noindent
\underbar{\sl Acknowledgements~} 
We thank the conveners
of the QCD/$\gamma\gamma$ working group for the
stimulating environment they have been able to create during the
workshops.

\end{document}